\documentclass{ws-procs9x6}

\usepackage{amsmath}
\usepackage{amssymb}

\begin{document}
\newcommand{\bean}{\begin{eqnarray}}
\newcommand{\eean}{\end{eqnarray}}
\newcommand{\eqs}[1]{Eqs. (\ref{#1})} 
\newcommand{\eq}[1]{Eq. (\ref{#1})} 
\newcommand{\meq}[1]{(\ref{#1})} 
\newcommand{\fig}[1]{Fig. \ref{#1}}

\title{Black holes and fundamental physics}

\author{Jos\'e P. S. Lemos}

\address{Centro Multidisciplinar de Astrof\'{\i}sica - CENTRA\\
Departamento de F\'{\i}sica, Instituto Superior T\'ecnico,\\
Universidade T\'ecnica de Lisboa,\\
Av.\ Rovisco Pais 1, 1049-001 Lisboa, Portugal.\\
E-mail: lemos@fisica.ist.utl.pt}

\maketitle

\abstracts{We give a review of classical, thermodynamic and quantum
properties of black holes relevant to fundamental physics.}
\noindent 
\footnote{Invited talk at the Fifth International Workshop 
on New Worlds in Astroparticle Physics, 
University of the Algarve, Faro, Portugal, January 8-10, 2005, 
published in the Proceedings of the
Fifth International Workshop on New Worlds in Astroparticle Physics, 
World Scientific (2006), eds. Ana M. Mour\~ao et al., p. 71-90. 
Talks also given at the Observat\'orio Nacional of Rio de Janeiro 
in November 2004, the University of Porto in December 2004, 
the Federal University of Rio de Janeiro in 
February 2005, the Classical University of Lisbon in April 2005, 
the University of Coimbra in June 2005, and the 
Technical University of Lisbon in July 2005.}

\section{Introduction}

The term black hole is usually associated with a large astrophysical
object that has formed due to huge gravitational fields that can arise
in the center of massive concentrations (see, e.g., 
\cite{lemosastroreview}).  However, the black hole is an
object in itself which should be studied within the domain of physics,
irrespective of the interactions with exterior astrophysical plasmas
which excite, and are excited by, the strong gravitational fields of
the black hole.  Here we want to understand a black hole as a physical
object.  This program was consciously initiated by Wheeler
\cite{wheeler1} back in the 1950s. We have not yet understood it
entirely, but we have come very far, if we think that, back in 1960,
Wheeler, Kruskal and others \cite{wheelerautobiog} 
managed to understand, for the first time, the global causal
structure of the complete manifold of the simpler black hole, the
Schwarzschild black hole. During the 1960s the black hole 
became well understood as a classical object, mainly due to 
the works of Penrose \cite{penrose}
and then Hawking \cite{hawking} and Carter \cite{carterleshouches}. 
But, then in the 1970s due to
Bekenstein \cite{bekenstein0} 
and Hawking \cite{hawking2}, the whole field
was revolutionized, the black hole concept entered the quantum
arena. Of course, quantum dynamics is the underlying dynamics of the
world and black holes have to be understood in this
context. Conversely, the simple structure of a black hole, can be used
to probe and learn about the quantum structure of gravitation.  As such, 
a black hole is considered by many \cite{bekenstein1} as the
gravitational equivalent of the hydrogen atom in mechanics, 
in the sense that this atom was
used by Bohr, Sommerfeld and others to touch and grasp the novel ideas
of quantum mechanics \cite{mehra}.

Hawking's monumental discovery of 1974, perhaps the most important
discovery in theoretical 
physics of the second half of the twentieth century, that a black
hole radiates quantum mechanically, was followed by some interesting
developments, but, perhaps, there was no ostensible growth, after
that.  Then, many physicists from different fields, like particle
physics and field theory, moved to string theory. String theory,
although a theory subject to criticisms on several grounds, can tackle
important problems related to black hole physics.  String theorists in
taking the black hole problem into their hands back in the 1990s
\cite{witten,horowitzstrominger}, out from the general relativists
alone, opened the subject to the physics community overall, and
revolutionized it into myriads of new directions. Before string theory
attacked the problem, one should obey the general relativity bible,
which was strict, allowing one to venture into other fields, such as
non asymptotically flat spacetimes or naked singularities, only with
extreme care and perhaps permission.  String theory opened up the book
and the theoretical discussion on black holes and their problems grew
exponentially.  Of course, general relativity itself benefited from
it.  For instance, new black hole solutions in general relativity, now
called toroidal black holes, living in asymptotically anti-de Sitter
spacetimes were found \cite{lemos1,2,3,4}, and many other connections
were made.

Thus, suddenly in the 1990s, the field of black holes was again lively
and growing fast.  Five topics of the utmost interest are: (a) the
thermodynamics of black holes, (b) the black hole entropy and its
degrees of freedom, (c) the information paradox, (d) the holographic
principle and its connection to the generalized second law and to the
covariant entropy bound, and (e) the inside of a black hole and its
singularity.  I will report on the first two topics, the others
require reviews on their owns, and will be left for other
opportunities.  Due to a large bibliography on these two first topics
I cannot be complete in listing references, the ones that are not
mentioned will be left to another larger review. I have benefited
tremendously from the reviews of Bekenstein
\cite{bekensteinreview1,bekensteinreview2} and Fursaev
\cite{fursaevreview}, see also the recent thorough reviews 
by Page \cite{pagereview} and Padmanabhan \cite{padmanabhanreview}.
Unless otherwise stated we use units in which
$G=c=\hbar=1$.

\section{Black hole thermodynamics and Hawking radiation}

\subsection{Preliminaries}

It is now certain that a black hole can form from the collapse 
of an old massive star, or from the collapse of a cluster of 
stars. Many x-ray sources observed in our galaxy contain a black 
hole of about 10$\,M_\odot$. Quasars, the most powerful distant 
objects, belong to a class 
which is one of the representatives of active galactic nuclei 
that are powered by a very massive black hole, with masses 
as high as $10^{10}M_\odot$. Our own Galaxy harbors a dead quasar 
with a mass of $10^{6}M_\odot$ in its core. 
Mini black holes, with masses lying within a wide range, 
a typical one could have $10^{15}\, {\rm gm}\sim 10^{-18} M_\odot$ 
(with radius $10^{-13}\,{\rm cm}$), 
may have formed in the early universe. 
Finally, Planck black holes of mass $10^{-5}\,{\rm gm}$ and radius 
$10^{-33}\,{\rm cm}$ may form in an astronomical collider 
which could provide a center of mass energy of $10^{19}\,{\rm GeV}$, 
or perhaps less if the idea of extra large dimensions is correct 
\cite{largeextradimensions}. 

A black hole is a gravitational object whose interior region
is invisible for the outside spacetime world. The 
boundary of this region is the event horizon of the black hole. 
To the outside world the black hole is like a tear in the 
spacetime, which interacts with its environment 
by attracting and scattering particles and waves in 
its neighborhood. 

Black holes appear naturally, as exact solutions, in the theory of
general relativity.  The most simple is the Schwarzschild black hole
which has only one parameter, the mass, and has a spherical horizon. By
adding charge one obtains a black hole with more structure, the
Reissner-Nordstr\"om black hole \cite{mtw}.  The theory of black holes
received a tremendous boost after Kerr \cite{kerr} found that a
rotating black hole is also an exact solution of general relativity, a
totally unexpected result at the time, that continues to 
flabbergast many people up to now. 
The Kerr black hole provides
extra non-trivial dynamics to the spacetime, from which novel ideas
sprang. Kerr black holes with charge are called Kerr-Newman black
holes, a name also used to designate the whole family.  There are now
other important families of black holes, such as the family 
of anti-de Sitter spacetimes, with negative cosmological constant, whose
horizons have topologies other than spherical \cite{4}, or other
families in a variety of different theories of gravitation
\cite{lemosreview1997}.  The Kerr-Neman family was the first to be
thoroughly investigated classically.  Some important properties,
generically valid for other families, have been worked out in detail. 

First, the event horizon acts as a one way membrane.  Due to the
strong gravitational field near the black hole, the light cones of the
spacetime get tilted, so much so that their exterior boundary lies
tangent to the horizon. The horizon thus acts as a one way membrane,
i.e., no object, not even a light ray, that crosses it inwards can
ever cross it back outwards. As a result, any physical quantity, such
as energy, entropy or information, that is damped into the black hole 
remains permanently trapped inside, classically.

Second, a black hole has no hair. What does this mean?  For an
exterior observer, placed outside the horizon, the black hole forgets
everything that it has swallowed.  The black hole can have been formed
from baryons alone, or from leptons alone, or from both, or anything
else, the exterior observer cannot have access to what formed the
black hole. The only thing the observer probes is the black hole mass
$M$, electromagnetic charge $Q$, and angular momentum $J$. This is
referred to as the baldness of the black hole, or as a black hole has no
hair, in the language of Wheeler.  In fact, it has three hairs,
$M,Q,J$, but the nomenclature is still correct, one usually associates
to someone that has three hairs that he is bald, has no hair!

Third, a black hole absorbs and scatters particles and waves.  These
properties involving scattering and absorption of particles and waves
by black holes, specially by rotating black holes, were very important
in the later developments. The whole subject started with the Penrose
process \cite{penrose}, which branched into superradiance on one hand and the
irreducible area concept on the other, 
and culminated with Hawking's theorem on
area growth. Let us comment on these features briefly, 
first with special
emphasis in superradiation.  When a particle is scattered by a Kerr
black hole and broken in two pieces in the process, energy can be
extracted from the black hole rotation into the outgoing particle,
using the existence of an ergosphere (a region just outside the event
horizon) \cite{penrose}, a kind of relativistic sling shot
phenomenom.  The wave analog of the Penrose process, whereby an
incoming wave (scalar, electromagnetic, gravitational or plasma) with
positive energy that impinges on the rotating black hole splits up
into an absorbed wave with negative energy and a reflected wave with
enhanced positive energy \cite{zeldovich1,unruh,bekenstein3}, is
called superradiance.  Consider a wave of the form $e^{-i\omega t + i
m \phi}$, where $\omega$ is the frequency of the wave, $t$ the time
parameter, $m$ the azimuthal angular wavenumber around the axis of
rotation of the black hole, and $\phi$ the angular
coordinate. Considering then that such a wave collides with the black hole,
one concludes that if the frequency $\omega$ of the incident wave
satisfies the superradiant condition $\omega < m \Omega\,,$ where
$\Omega$ is the angular velocity of the black hole, then the scattered
wave is amplified \cite{zeldovich1,unruh,bekenstein3},
One simple way to get an idea of what is happening is by resorting to
the inverse of the characteristic frequencies, i.e, the period of the
wave $\tau=2\pi/\omega$, and the rotation period of the cylinder
$T=2\pi/\Omega$. Then the superradiant condition is now $m\,\tau>T$,
which means that for $m=1$ say, the wave suffers superradiant
scattering if it takes a longer time in the neighborhood of the
black hole than the time the black hole takes to make one revolution, so
that there is enough time for the black hole to transfer part of its
rotating energy to the wave.  This way of seeing superradiance
corresponds to giving a necessary condition, i.e., to exist
superradiance there should exist enough time so that the black hole
can transmit part of its energy to the wave.

Fourth, the area of a black hole always grows in any physical
process. The Penrose process also led to the concept of
irreducible mass \cite{Christod}, which in turn led Hawking
\cite{HawkingPRL} to prove a theorem stating that the black hole area
always grows in any physical process, classically. This theorem proved
to be decisive for further developments. In turn, 
and in passing, one can use this
theorem to prove superradiance.  Indeed, following the lines of
Zel'dovich \cite{zeldovich1} one roughly finds that the scattering of a
wave by a black hole obeys 
$\frac{\kappa}{8\pi}\,\frac{dA}{dt}=\left( P_i-P_r \right ) \left (
1-\frac{m\Omega}{\omega} \right )\,,
\label{heuristicbh}$
where $A$ is the area of the event
horizon, $\kappa$ is its surface gravity, and $P_i$ and $P_r$ 
are the incident and reflected power of the wave, respectively. 
From the area law for black holes, which states that
the area of the event horizon never decreases, i.e, $d\,A\geq0$
\cite{HawkingPRL}, one finds that if the frequency of the incident
wave satisfies the superradiant condition, the second factor in the
right hand side of the equation is negative. In order to
guarantee that the area does not decrease during the scattering
process, one must have $P_r>P_i$. Thus, the energy
of the wave that is reflected is higher than the energy of the
incident wave, as long has the superradiant condition is
satisfied. On other developments on superradiance and how it can be used,
along with a mirror, to build a black hole bomb see 
\cite{blackbomb,gulbenk2004}.

With these four ingredients, i.e., one-way membrane, no hair,
scattering properties, and area law, all is set to put the black
hole in a thermodynamic context.

\subsection{Thermodynamics and Hawking radiation}

A Kerr-Newman black hole, say, can form from the collapse of 
an extremely complex
distribution of ions, electrons and radiation. But once it has formed
the only parameters we need to specify the system are the parameters
that characterize the Kerr-Newman black holes, the mass $M$, the
charge $Q$ and the angular momentum $J$.  Thus we have a system
specified by three parameters only, which hide lots of other
parameters.  In physics there is another instance of this kind of
situation, whereby a system is specified and usefully described by few
parameters, but on a closer look there are many more other parameters
that are not accounted for in the compact description. This is the
case in thermodynamics.  For thermodynamical systems one gives the
energy $E$, the volume $V$, and the number of particles $N$, say, and
one can describe the system in a usefully manner, although the system
encloses, and the description hides, a huge number of molecules.

Connected to this, was the question Wheeler was raising in the
corridors of Princeton University \cite{wheelerautobiog}, that in the
vicinity of a black hole entropy can be dumped onto it, thus
disappearing from the outside world, and grossly violating the second
law of thermodynamics.  Bekenstein, a Ph.D. student in Princeton at
the time, solved part of the problem in one stroke.  He postulated,
entropy is area, more precisely \cite{bekenstein0}, 
$S_{\rm BH}=\eta\,\frac{A}{l_{\rm pl}^2}\;\,k_{\rm B}\,, 
\label{entropybekenstein}$
where one is using full units, 
$\eta$ is a number of the order of unity or so, that could not
be determined, $l_{\rm pl}\equiv\sqrt{\frac{G\hbar}{c^2}}$ is the
Planck length, of the order of $10^{-33}\,$cm, and $k_{\rm B}$ is the
Bolztmann constant.  This is, of course, aligned with the area's law of
Hawking, and the Penrose and superradiance processes.  Bekenstein
invoked several physical arguments to why the entropy $S$ should go
with $A$ and not with $\sqrt A$ or $A^2$.  For instance, it cannot go
with $\sqrt A$ ($A$ itself goes with $\sim M^2$) 
because when two black holes merge the final mass obeys
$M<M_1+M_2$ since there is emission of gravitational radiation. 
But if $S_{\rm
BH}\propto M<M_1+M_2\propto S_{\rm BH1}+S_{\rm BH2}$ the entropy could
decrease, so such a law is no good. The correct option turns out to
be $S\propto A$, the one that Bekenstein took. Also correct, it seems,
is to understand that this is a manifestation of quantum gravity, so
that one should divide the area by the Planck area, and multiply by
the Boltzmann constant to convert from the usual area units into the
usual entropy units.  There is thus a link between black holes and
thermodynamics.

One can then wonder whether there is a relation obeyed by black hole
dynamics equivalent to the first law of thermodynamics. For a
Schwarzschild black hole one has that the area of the event horizon
is given by $A=4\pi\,r_+^2$. Since $r_+=2M$ one has $A=16\pi M^2$.
Then one finds $dM=1/(32\,\pi\,M)\, dA$, which can be written as. 
\begin{equation}
dM=\frac{\kappa}{8\,\pi} dA\,,
\label{firstlawofbhs}
\end{equation}
which is the first law of black hole dynamics \cite{bardeenetal}.
The surface gravity of
the event horizon of the Schwarzschild black hole 
is $\kappa=1/4\,M$.  
Equation (\ref{firstlawofbhs}) can be compared with
\begin{equation}
dE=TdS\,,
\label{firstlawofthermodynamics}
\end{equation}
which is the first law of thermodynamics. Note that, a priori, the
analogy between $S$ and $A$, and $T$ and $\kappa$, is merely
mathematical, whereas the analogy between $E$ and $M$, is physical,
they are the same quantity \cite{waldbook}. For a generic Kerr-Newman
black hole one has the relation $dM=\frac{\kappa}{8\pi}\, dA+\Omega\, dJ+
\Phi\, dQ\,,\label{firstlawofbhs2}$ where 
$\Omega$ is the angular velocity of the black hole
horizon, and $\Phi$  the electric
potential. Comparing with the thermodynamical relation 
$dE=T\,dS+p\,dV+\mu\,dN\,,\label{firstlawofthermodynamics2}$ 
where the symbols have their
usual meanings, it further strengthens the analogy.

Following Bekenstein, this is no mere analogy though, the black hole system
is indeed a thermodynamic system with the entropy of this system being
proportional to the area. But what is $\eta$ in the equation
proposed by Bekenstein? Thermodynamic arguments alone were not
sufficient to determine this number.  Using quantum field theory
methods in curved spacetime Hawking \cite{hawking2} showed that a
Schwarzschild black hole radiates quantically as a black body at
temperature
\begin{equation}
T=\frac{1}{8\pi M}\,.
\label{hawkingtemperature}
\end{equation}
Since $\kappa=1/4\,M$, the temperature and the surface gravity are
essentially the same physical quantity, with $T=\kappa/2\,\pi$. Moreover,
from equation (\ref{firstlawofthermodynamics}), one obtains
$\eta=1/4$, yielding finally 
\begin{equation} S=\frac14\,A\,.
\label{entropyarealawgeometrical} 
\end{equation} 
in geometrical
units. Thus the Hawking radiation solved definitely the thermodynamic
conundrum. However, it introduced several others puzzles.

The black hole is then a thermodynamic system.  Thus, the second law
of thermodynamics $\Delta\,S\geq0$ should be obeyed. Since one does
not know for sure the meaning of black hole entropy, it is useful to
write the entropy as a sum of the black hole entropy $S_{\rm BH}$,
and the usual matter entropy $S_{\rm matter}$, i.e., $S=S_{\rm
BH}+S_{\rm matter}$, allowing one to write the second law as
\begin{equation}
\Delta\,S_{\rm BH}+\Delta S_{\rm matter}\geq0\,,
\label{gsl}
\end{equation}
commonly called the generalized second law 
\cite{bekensteingeneralizedsecondlaw}. 
The generalized second law proved important in many developments.

\section{Statistical interpretation of black hole entropy} 
\label{statistical} 

\subsection{Preliminaries}

In statistical mechanics, the entropy of an ordinary object is a
measure of the number of states available to it, i.e., it
is the logarithm of the number of quantum states that the object may 
access given its energy. This is the statistical meaning of the
entropy. Since black holes have entropy, 
one can ask what does the black hole entropy represent?  What is the
statistical mechanics of a black hole as a thermodynamic object?

Retrieving full units to equation (\ref{entropyarealawgeometrical}) 
one has 
\begin{equation}
S_{\rm BH}=\frac14\,\frac{A}{l_{\rm pl}^2}\;\,k_{\rm B}\,,
\label{bhentropyfullunits}
\end{equation}
where again, $l_{\rm pl}^2=\frac{c^2}{\hbar\,G}$ is the Planck area, 
and $k_{\rm B}$ is the Boltzmann constant. Judging from 
the four fundamental constants appearing in the formula, namely, 
$G,c,\hbar,k_{\rm B}$, one gets a system where 
relativistic gravitation, quantum mechanics and 
thermodynamics, are mixed together, 
indicating that a statistical interpretation should be in sight. 
Moreover, the number $\frac{A}{l_{\rm pl}^2}$ itself suggests 
that the sates of the black hole are some kind or another 
of quantum states. In addition, 
the factor $1/4$ became a target for 
any theory that wants to explain black hole entropy from 
fundamental principles.

Note that black hole entropy is large. A neutron star with one solar
mass has entropy of the order of $S\sim10^{57}$ (in units where
$k_{\rm B}=1$) in a region within a radius of about 10$\,$Km. A solar
mass black hole has an entropy of $10^{79}$ in a region within a radius
of 3$\,$Km.  There is a huge difference in entropy for these two
objects of about the same size, suggesting, somehow, 
the black hole harnesses
entropy that can be peeled away through the black hole's lifetime,
i.e., the time the black hole takes to radiate its own mass via
Hawking radiation.

\subsection{Entropy in the volume}

Bekenstein \cite{bekenstein75} tried first to connect the entropy of a
black hole with the logarithm of the number of quantum configurations
of any matter that could have served as the black hole origin, in
perfect consonance with the no hair theorem. Now, the number of those
quantum configurations can be associated to the number of internal
states that a black hole can have, hinting in this way that the
entropy of a black hole lies on the volume inside the black hole
(see also \cite{frolovnovikov}). This
idea of bulk entropy, although interesting, has many drawbacks, see
\cite{jacobson,jacobsonmarolf,sorkinlast}.

\subsection{Entropy in the area}

There are now many alternative interpretations that associate the 
black hole entropy with the its area, the area of the 
horizon. One can divide these interpretations into those 
that claim the degrees of freedom are on the quantum matter 
in the neighborhood of the horizon that gives 
rise to the Hawking radiation, those that claim that 
the degrees of freedom are on the gravitational field alone, 
and those that put the degrees of freedom on both, matter 
and gravitational fields, like string theory.

\subsubsection{Matter entropy}

{\it (i) Entropy of quantum fields}
\vskip 0.1cm
\noindent  One interpretation says that the black hole entropy
comes from the entropy of quantum matter fields fluctuations in the
vicinity of the horizon. This was first advanced by Gerlach 
\cite{gerlach} who proposed that the entropy was related to the number
of zero-point fluctuations that give rise to the Hawking radiation. So
the entropy comes from the all time matter fields surrounding the horizon
created by the Hawking process. Later Zurek and Thorne \cite{zurek}
proposed the quantum atmosphere picture and 't Hooft \cite{thooft}
developed the idea in the brick wall model. The advantage of these
insights is that the linear dependence of $S_{\rm BH}$ on the horizon
area, $S_{\rm BH}=\eta A$ comes automatically, since the matter that
gives rise to the entropy is in a thin shell surrounding a surface,
the horizon. One great disadvantage, is that the coefficient $\eta$ is
infinite since ultraviolet wavelengths, with wavelengths arbitrarily
small, also take part in the matter fields surrounding the horizon.
One can cure this by imposing a cutoff for these lengths at the
Planck length, which, although sensible, is ad hock and incapable of
giving the goal factor 1/4. Moreover, $\eta$ is proportional to the
number of fields existing in nature, making even harder to connect it
with the coefficient 1/4.

\vskip 0.3cm
\noindent {\it (ii) Entropy of entanglement}
\vskip 0.1cm
\noindent Another, somehow connected, 
interpretation comes from Sorkin and collaborators 
\cite{sorkin86}, who suggested that the entropy is related to the entanglement 
entropy arising from tracing out the degrees of freedom existing beyond 
the horizon. In other words, the entropy is generated by dynamical degrees 
of freedom, excited at a certain time, associated to the matter in the 
black hole interior near the horizon through non-causal EPR correlations 
with the external matter. It has been used by many different authors, 
see e.g., \cite{srednicki,adami}.  
This has the 
advantages and disadvantages of the above interpretation.  

\vskip 0.3cm
\noindent {\it (iii)  Entropy in induced gravity}
\vskip 0.1cm
\noindent 
Other interpretation that can be mentioned is the one that 
associates the degrees of freedom to heavy matter fields 
that when integrated out induce, naturally, general relativity. 
This way of seeing general relativity was envisaged by 
Sakharov\,\cite{sakharov} and the corresponding entropy 
interpertation was put forward in\,\cite{frolovfursaev}. 
Of course, this interpreation could be in the  
gravitational entropy sector, since what is matter and what is gravity 
is blurred here.

\subsubsection{Gravitational entropy}

\noindent {\it (i) Entropy from boundary conditions}
\vskip 0.0cm
\noindent 
An improved interpretation, perhaps, is that of 
Solodukhin\,\cite{solodukhin} and 
Carlip\,\cite{carlip99,carlip02,carlip05} 
who, independently, switched from matter field fluctuations 
to gravitational field fluctuations.
They showed  
that the existence of a horizon, the surface where the 
fluctuations occur, makes the fluctuations themselves obey the laws of a 
conformal field theory in two spatial dimensions, this number two is 
related to the dimensionality of the horizon. Conformal field theory  
has been thoroughly investigated, yielding for the logarithm of 
number of states 
associated with the fluctuations, a value for the entropy that matches exactly 
the entropy formula for a black hole, with the coefficient 1/4 coming out 
perfect. The idea is to use the correct boundary conditions at a horizon 
so as to give rise to new degrees of freedom that do not exist in the 
bulk spacetime. However interesting it may be, see also \cite{gdiaslemos}, 
it lacks a direct physical interpretation, since the boundary 
conditions are too 
formal.

\vskip 0.1cm
\noindent {\it (ii) Heuristic interpretation for the degrees of freedom}
\vskip 0.0cm
\noindent 
A physical interpretation for the gravitational degrees of freedom 
comes from the intuitive idea of Bekenstein and Mukhanov 
\cite{bekmukha,bekenstein1} 
that the area of the horizon being an adiabatic 
invariant, should be quantized in Ehrenfest's way.  
Suppose, then, that the area of the horizon is quantized 
with uniformly spaced levels of order of the Planck length squared, i.e., 
$A=\alpha\,l_{\rm pl}^2\,n$ with $\alpha$ a pure number, and 
$n=1,2,....\,$. Thus a small black hole is constructed from a small number 
of Planck areas, one can build the next black hole putting an extra 
Planck area, and so on. The horizon, according with this view, 
can be thought of as a patchwork of patches with area $\alpha\,l_{\rm pl}^2$. 
If every Planck patch can have two distinct states, say, then a black 
hole with two Planck areas can be in four different states, 
a black hole with three Planck areas can be in eight different states, 
a large black hole with $n$ Planck areas can be in $2^n$ 
different surface states. Now, degeneracy and entropy are connected 
in such a way that latter is the logarithm of the former, i.e., 
$S_{\rm BH}= \ln\,2^n=(\ln2)\,n=\frac{\ln2}{\alpha}\frac{A}{l_{\rm pl}^2}$. 
The area law is then recovered, by default. Further, 
from Hawking's work we know that $\frac{\ln2}{\alpha}=\frac14$ so that the 
quantization law is $A=4\,(\ln2)\,l_{\rm pl}^2\, n$. We can instead think 
that every area patch has $k$ distinct states instead of two. Then the 
same reasoning follows, and one has that a black hole with area 
$A=\alpha\,l_{\rm pl}^2\,n$ can be in any of $k^n$ sates. The entropy 
is then $S_{\rm BH}= \ln\,k^n=(\ln k)\,n=\frac{\ln2}{\alpha}
\frac{A}{l_{\rm pl}^2}$, and the area quantization law is 
$A=4\,(\ln k)\,l_{\rm pl}^2\, n$, and $\alpha=4\ln k$. 
The question is now, what is $k$? Hod \cite{hod} found a way to 
determine $k$. Inspired by Bohr's correspondence principle, 
that transition frequencies at large quantum numbers should equal 
classical oscillation frequencies, one should associate the classical 
oscillation frequencies of the black hole 
with the highly damped quasinormal frequencies, since these take no time, 
as quantum transitions take no time. So, for instance, the highly 
damped quasinormal frequencies of the Schwarzschild black hole are found 
to be $M\,\omega_n=\frac{\ln3}{8\pi}-
\frac{i}{4}\left( n+\frac12\right)$, to leading order. The factor 
$\frac{\ln3}{8\pi}$ was first found numerically \cite{who?}, and much later 
analytically \cite{motleneitzke}. Then using $\Delta M=\omega$ and so 
$\Delta A=32\,\pi\,M\,\Delta M=32\,\pi\,M\,\omega=4\,(\ln3)\, l_{\rm pl}^2$, 
along with, from the very definition of $A$,  
$\Delta A=4\,(\ln k)\,l_{\rm pl}^2 
\Delta n$, constrained by $\Delta n=1$ 
as it is required for a single simple area 
transition, one finds $k=3$. Then the quantization of the area 
is given by $A_n=4\left(\ln3\right)\, l_{\rm pl}^2\, n$. 
This has been also used by people of loop quantum gravity, and received 
a boost as the whole idea of Hod fixes the Barbiero-Imirzi 
parameter, a loose parameter in the theory \cite{baez}.

The spin-area parameter $k$ was fixed in the case of 
a Schwarzschild black hole, $k=3$. What can one say about the other 
black holes? 
The subject of quasinormal modes is a subject in which research has 
been very active since Visheveshwara noticed 
that the signal from a perturbed black hole is, for most of the 
time, an exponentially decaying ringing signal, with the 
ringing frequency and damping timescales being characteristic of the 
black hole, depending only on its parameters like $M$, $Q$ and $J$, 
and the cosmological constant $\Lambda$, say. 
Whereas for astrophysical black holes the most 
important quasinormal frequencies 
are the lowest ones, i.e., frequencies with small imaginary part, 
so that the signal can be detected, for black holes 
in fundamental physics 
the most important are 
the highly damped ones, since one is interested in the 
transition between classical and  the quantum physics 
(see, e.g., \cite{cardosolemos} and for a review \cite{cardosothesis}).
Ultimately, one wants to understand whether the number $k=3$ 
depends on the nature of the black 
hole (does a Kerr black hole give the Schwarzschild number), 
on the nature of spacetime (asymptotically flat, 
de Sitter, anti-de Sitter), and on the dimension
of spacetime or not. 
Different spacetimes yield different 
boundary conditions, and thus completely different behavior 
for quasinormal 
modes, whereas one might expect black hole area levels to depend only 
on local physics near the horizon, so that 
it is not obvious how to 
reconcile such locality with the quasinormal mode behavior. 
This makes it hard to argue that $k$ is universal, as it should be.
The study on other different black holes has not been conclusive.

\vskip 0.3cm
\noindent {\it (iii) York's interpretation }
\vskip 0.1cm
\noindent 
York \cite{york} made a very interesting proposal where the entropy of
the black hole comes from the statistical mechanics of zero-point
quantum fluctuations of the metric, in the form of quasinormal modes,
over the entire time of evaporation. 
The approach has thus a very physical 
interpretation for the entropy, and gets the coefficients for 
the black hole entropy and temperature within 
the same order of magnitude as the exact ones. 
York's idea is the translation of
Gerlach's quantum matter fields fluctuations \cite{gerlach} to 
fluctuations in the gravitational field, and has been retaken
in \cite{loweetal}.

\vskip 0.3cm
\noindent {\it (iv) Other methods}
\vskip 0.1cm
\noindent 
Other methods are Euclidean path integral \cite{gibbonshawking}, 
giving $S_{\rm BH}=\frac14 A$ directly, but it is flawed, 
since it uses a saddle
point approximation at a point that is not a minimum. There is a
method of surface fields and Euclidean conical singularities
\cite{teitel1996}.  There is the Noether's charge method \cite{wald},
a very useful one that has been frequently used.  There are also hints
that the entropy depends on the gravitational Einstein-Hilbert action
alone, and like energy in general relativity, is a global concept
\cite{hawkinghunter1999}.

There are other techniques that, although not constructed 
to yield an interpretation, 
corroborate that there should be a statistical interpretation. 
One of these is related to pair creation 
of black holes. In the Schwinger process of production of charged 
particles in a background electric field, the total production rate 
grows as the number of particle species produced. If this is 
extrapolated to black hole production in a background field then 
the rate of the number  of black hole pairs produced should go 
as the number of black hole states. Indeed, one can show 
that the factor $\Gamma\sim {\rm e}^{\frac14 A_{\rm BH}}=e^{S_{\rm BH}}$ 
multiplies indeed the pair 
production amplitude, consistent with interpretation that the 
entropy counts black hole microstates. To work out these results one has 
to find the instanton solution, i.e., the solution that gives 
the transition rates, of the Euclidean C-metric, where the C-Metric 
is the solution for two black holes accelerating apart. This 
has been done for asymptotically flat spacetimes \cite{garfinkleetal}
and for de Sitter and anti-de Sitter spacetime 
\cite{diaslemos0,diaslemos1,diaslemos2,diasthesis}. 

The notion of black hole entropy has been extended to higher
dimensional spacetimes where one can also have black $p$-branes. A
black hole is a special case of a black $p$-brane, one with $p=0$, a black
string has $p=1$, a black membrane has $p=2$, and so on.
These black branes suffer from a gravitational instability, 
the Gregory-Laflamme instability, and  entropic arguments suggest that 
the fate of such a brane is a set of black holes \cite{kol}.

\subsubsection{Entropy in string theory}

So far, I have not mentioned what is the contribution of string theory
to the interpretation of black hole entropy. String theory has been
extremely helpful in the advancement of black hole theory for several
reasons. In relation to the calculation of black hole entropy it has
given new methods, and many new and different black hole solutions on
which one can apply these new methods. On the other hand, in relation
to the interpretation of black hole entropy, to answer the question of
where are the degrees of freedom, it has come short of a result. Let
us see several developments in the context provided by string theory. 

\vskip 0.3cm
\noindent {\it (i) Heuristics }
\vskip 0.1cm
\noindent First, heuristics \cite{horowitzpolchinski}. 
String theory is a theory that provides many fields, which 
can be called matter fields, besides the gravitational field, 
and so the degrees of freedom for the entropy can come from 
both, the matter and the gravitational fields, now studied 
together in a coherent fashion.
Since a string 
is matter, the entropy of a string goes with mass, and one 
can write $S_{\rm string}\simeq l_{\rm s}\,M\,$, where 
$l_{\rm s}$ is the fundamental string length 
(i.e., the string lengthscale), 
and $M$ is the mass of the string 
(here we use string units). The entropy of a Schwarzschild 
black hole goes as $S_{\rm BH}\simeq G\,M^2\simeq g^2\,l_{\rm s}^2\,M^2$, 
where now it was advisable to recover $G$, which in string units is 
equal to $g$ the coupling of the string with the spacetime times 
the string lengthscale $l_{\rm s}$, both squared. 
The black hole radius goes as 
$r_{\rm BH}\simeq G\,M\simeq g^2 l_{\rm s}^2\,M$. Now, the coupling 
$g$ can be changed. Start from a black hole state, and assume one 
decreases the coupling reversibly and adiabatically, i.e., maintaining 
$S_{\rm BH}$ constant. Then $M\sim 1/g$ increases, 
and $r_{\rm BH}\sim g$ decreases. Thus as one puts less coupling,
maintaining the entropy,  
the mass of the black hole increases so as to compensate in the 
number of states; on the other hand the radius decreases because 
there is much less gravity, a behavior that is similar 
to polytropic white dwarf stars. 
Now, one cannot go on decreasing the radius forever, the process 
has to stop when the radius of the black hole is of the order of 
the string scale $r_{\rm BH}\sim l_{\rm s}$. So, 
$l_{\rm s}\sim g_{\rm crit}^2\,l_{\rm s}^2\,M_{\rm crit}$ yielding 
from the black hole side 
$M_{\rm crit}\sim1/(g_{\rm crit}^2\,l_{\rm s})$. This, in turn, 
implies $S_{\rm BH}\sim1/g_{\rm crit}^2$, and from 
the string side $S_{\rm string}\sim1/g_{\rm crit}^2$. 
Thus, this heuristic reasoning gives that there is a transition point
from the black hole state to the string state, and vice versa, meaning
that heavy string states form black holes, a not unexpected result.
Unfortunately, there is no control as to where are the degrees of
freedom when the black hole forms in this set up.  One knows where are
the degrees of freedom of the string, in the string itself, the way it
curves, wiggles, vibrates, and so on, but then when it collapses and
turns into a black hole at the transition point, it is a usual
gravitational collapse, leaving us again in the dark.  The nice thing
about this calculation is that at the transition point the entropy is
about the same for string and black hole, but how is the entropy
transfered from the string to the black hole, or vice-versa, the
calculation leaves us blind. See, however, the fuzzball 
proposal for black holes \cite{mathur}, where there is 
a retrieval  of the intepretation that 
the entropy of a black hole lies on the volume inside the black hole, 
not in its area.

\vskip 0.3cm
\noindent {\it (ii)  Exact calculations for extreme black holes }
\vskip 0.1cm
\noindent
Extreme black holes allow an exact, though tricky, calculation of 
the entropy, done for the first time by 
Strominger and Vafa  \cite{stromingervafa}. 
A simple extreme black hole has mass $M$ and charge $Q$ that 
obey the relation $Q=(\sqrt G)M$. The entropy is then 
$S_{\rm BH}=4\,\pi\, G\,M^2=4\,\pi\,Q^2$. 
Since the entropy does not depend on the 
gravitational constant $G$, the entropy is a measure of the number of 
the elementary charges of the extreme black hole alone. 
As we now know, $G=g^2 l_{\rm s}^2$, and so the entropy does not depend 
on $g$. One can vary the string coupling $g$ and obtain the same entropy.
On the other hand, $r_{\rm BH}=G\,M=\sqrt G\, M=g\,l_{\rm s}\,Q$, 
so $r_{\rm BH}$ depends on $g$. For weak coupling one has 
$g<<1$ and so $r_{\rm BH}<<l_{\rm s}$, the object is a condensed string 
in an almost flat spacetime,  
it is actually an intricate 
condensate of strings and branes, whereas for strong coupling one has 
$g>>1$ and so $r_{\rm BH}>>l_{\rm s}$, the object is a black hole. 
Now, some extreme black holes have the property they are supersymmetric, 
i.e., supersymmetric transformations do not change the black hole, and 
there are some theorems that say that there are no quantum corrections 
when going from strong to weak coupling and vice-versa. So, one 
can calculate the entropy of the object at weak coupling, where one 
has an object in flat spacetime and then extrapolate directly 
and exactly this calculation to strong coupling. At weak coupling, 
one finds that the dual theory that governs the dynamics of the condensate 
of branes and strings is a conformal field theory. One can then use 
the machinery of conformal field theory, through the Cardy formula, 
and get the entropy. Amazingly, for certain black holes in 
string theory, with several 
different charges, it gives exactly the black hole entropy. 
This calculation is very interesting indeed, but again it 
leaves us blind to what are and where are the black hole degrees of freedom.
Another snag of the calculation, is that it does not work out for 
general black holes, it works out only for extreme black holes, and 
even so not all extreme black holes.

\vskip 0.3cm
\noindent {\it (iii)  What is conformal field theory? }
\vskip 0.1cm
\noindent
We have been talking about 
conformal field theory, in various connections, namely in connection 
with the degrees of freedom of the horizon related to the method 
of Carlip \cite{carlip99} 
and Solodukhin \cite{solodukhin}, and in connection with the string theory 
methods. But what is conformal field theory? A way to see this 
\cite{fursaevreview}
is to work with $c$ massless scalar fields in one spatial dimension, 
i.e., in two spacetime dimensions. The Klein-Gordon equation for 
each field is 
\begin{equation}
\left(\partial_t^2-\partial_x^2\right)\phi_k(t,x)=0\;,\quad k=1,...,c\,,
\label{kleingordon}
\end{equation}
valid in a one dimensional box of length $b$, i.e, 
with boundary conditions 
given by $\phi_k(t,0)=\phi_k(t,b)=0$. This has a 
Planck radiation spectrum whose free energy is 
given by 
$
F(T,b)=c\,T\;\Sigma_n
\ln\left(1-{\rm e}^{-\omega_n/T}\right)\,,\, \omega_n=\frac{\pi}{b}\, n\,,
n=1,2,...,
\label{planckspectrum}
$
where the $\omega_n$ are the normal frequencies of the fields. 
In the thermodynamic limit 
$T\,b>>1$, one can evaluate the sum to obtain $F(T,b)=-\frac{\pi c}{6}
\, b\, T^2$, and thus 
\begin{equation}
S(E,b)=2\pi\sqrt{\frac{c}{6}\frac{b\,E}{\pi}}\,.
\label{entropythermodynamic}
\end{equation}

Now, how can one calculate this entropy using conformal field theory 
methods. First, one notes that the theory given in equation 
(\ref{kleingordon}) is indeed conformal invariant. 
Using null coordinates $x_-=t-x$ and $x_+=t+x$ 
the Klein-Gordon equation turns into $\partial_{x_-}\partial_{x_+}\phi_k=0$. 
Indeed, this is invariant under conformal transformation $x_-\rightarrow
x_-'=f(x_-)$ and $x_+\rightarrow x_+'=f(x_+)$. 
Now, when one has a symmetry, in this case conformal, one has an associated 
conserved charge. In turn these conserved charges are the generators 
of the corresponding symmetry transformation (for instance, the 
Hamiltonian is the generator of time translations, translations 
being included in conformal transformations).
In two dimensions the generators, $L_n$ and $\bar{L}_n$, of conformal 
transformations are infinite. They give the 
standard Virasoro algebra, $[L_n,L_m]=(n-m)\,L_{n+m}$, and the 
same for the complex conjugate, where the brackets are Poisson brackets. 
Interesting to note that the algebra of the generators is the same as 
the algebra of the Fourier components of the infinitesimal vector 
field that gives the coordinate transformations. 
The Hamiltonian generator is $\frac{2\pi}{b}
(L_0+\bar{L}_0)$. This is classical, and 
there is no entropy for the $c$ scalar fields. 
However, when quantized the generators get an extra term, 
quantum mechanics yields always a scale which in turn 
produces an anomaly in 
the conformal field theory. This 
gives rise to an extra term for the algebra, 
\begin{equation}
[L_n,L_m]=(n-m)\,L_{n+m}+\frac{c}{12}(n^3-n)\,\delta_{{n+m}\;0}\,, 
\label{virasoroalgebra}
\end{equation}
where the brackets should be viewed as a commutator now, and 
the generators as operators. 
The Hamiltonian operator is then $H=\frac{2\pi}{b} (L_0+\bar{L}_0)$, 
which when applied in a state $|h,\bar h>$ gives the energy 
$E=\frac{2\pi}{b}(h+\bar h)$. Now, the 
sate $|h,\bar h>$ can be constructed from vacuum, 
the box without fields, in many different ways, since 
$|h,\bar h>=\Pi_k(L_{-k})^{\alpha_k}\,\Pi_p(\bar{L}_{-p})^{\beta_p}|0>$, 
with $\Sigma_k\,\alpha_k=h$, $\Sigma_p\,\alpha_p=\bar h$, 
and where $|0>$ is the vacuum vector, and $L_{-k}$ are creation operators. 
The state $|h,\bar h>$ is an eigenstate of $L_0$
and $\bar{L}_0$, sure. One can then find the degeneracy, as Cardy did 
\cite{cardy}, 
and show that $D={\rm e}^{2\pi\sqrt{\frac{c}{6}\frac{bE}{\pi}}}$. 
Thus the entropy of the $c$ conformal fields in a periodic box is $S=\ln\,D
=2\pi\sqrt{\frac{c}{6}\frac{bE}{\pi}}$, as in the thermodynamic result. 

In possession of these ideas, we can better understand 
the Strominger-Vafa calculation. For low $g$ one has a condensate 
of strings  and branes instead 
of a black hole, which obey a conformal field 
theory. With the theory in hand one finds $c$, $E$ and $b$, then one gets 
$S_{\rm CFT}$, and through supersymmetry arguments, extrapolates 
to high $g$, giving $S_{\rm BH}$, through 
$S_{\rm BH}=S_{\rm CFT}$. The entropy $S_{\rm BH}$ obtained 
in this way gives precisely $S_{\rm BH}=\frac14 A$. This is exact, 
but no interpretation for the entropy.

\vskip 0.3cm
\noindent {\it (iv)  The BTZ black hole and the AdS/CFT conjecture}
\vskip 0.1cm
\noindent
There is another place where these calculations are exact, it is the
three dimensional BTZ black hole that lives in a cosmological
constant $\Lambda$ background, i.e., in an anti-de Sitter spacetime. 
The idea \cite{strominger}  came as follows. 
Brown and Henneaux \cite{brownhenneaux} 
showed for the first time that the asymptotic group of
three-dimensional anti-de Sitter spacetime is the conformal group
in two dimensions, stating in addition that any quantum theory of such
a type of spacetime should take this into account. At about the same
time Cardy gave a formula, now famous, for the entropy of a
two-dimensional conformal field theory with central charge 
\cite{cardy}.  Then, later, 
Strominger \cite{strominger} applied the Cardy formula to 
Brown and Henneaux 
results \cite{brownhenneaux} 
and showed that it gave the formula discovered by Bekenstein and 
Hawking, $S_{\rm BH}=\frac14\,A$ \cite{bekenstein0,hawking2}. 
More precisely, in
this spacetime one has an intrinsic length scale, which is
$l=1/\sqrt\Lambda$.  One also has the black hole radius 
$r_{\rm BH}$. Now, the black hole entropy can be calculated through
gravitational methods to give $S_{\rm BH}=\frac14 A= \frac14
\frac{2\pi\,r_{\rm BH}}{G}=2\pi\,\sqrt{\frac{l^2\,M}{2\,G}}$, with
$M=\frac{r_{\rm BH}^2}{8\,l^2\,G}$. Now, compare with the Cardy
formula $S_{\rm CFT}=2\pi\,\sqrt{\frac{c\,b\,E}{6\pi}}$. For this put
$b=2\pi\, l$, forcing the conformal field theory to live on a cylinder
of perimeter $2\pi\, l$, identify $M=E$, and then choose the central
charge as $c=\frac32\frac{l}{G}$ \cite{strominger}.  Then, with these
choices $S_{\rm BH}=S_{\rm CFT}$. This equation relates classical and
quantum quantities. The conformal theory is quantum lives on a flat
spacetime $M_2$, one dimensional lower than the black hole, which
lives in three dimensional spacetime. The metric on $M_2$ is a
cylindrical flat metric, $ds^2=-dt^2+l^2\,d\varphi^2$. On the other
hand, the metric for the black hole spacetime at constant large radius
is $ds^2=\frac{r^2}{l^2}(-dt^2+l^2\,d\varphi^2)$. So $M_2$ can be
seen, apart from a superfluous factor, as the asymptotic infinity of
$M_3$, as its asymptotic boundary.  Therefore, the black hole
entropy (a semiclassical limit of quantum gravity), is determined by a
quantum conformal field theory (CFT) defined at the asymptotic
infinity of the bulk anti-de Sitter (AdS) spacetime.  This is an
example of the AdS/CFT conjecture of Maldacena
\cite{maldacenaAdsCFTconjecture}, which was based in other
spacetimes, and also works here. 
Since this type of computation for the black hole 
entropy is done at infinity, the infinity of anti-de Sitter 
spacetime, it does not see the details of the 
horizon. Thus, more than a direct computation of black 
hole entropy, this type of computation gives an upper bound 
for the entropy of anti-de Sitter spacetime in three dimensions. 
In this case, it is just as good, since the maximum of entropy in 
a region arises by inserting a black hole in it.

\section{Conclusions}

Thus we see that we are still far from having a consensus
\cite{jacobsonmarolf}.  Are the degrees of freedom located in the
volume or in the area, or in both, or are they complementary
descriptions?  Are they realized in the matter or in the gravitational
field or in both?  The answer still lies ahead.  
The entropy puzzle
does not exhaust the black hole.  Other sources of fascinating
problems and conundrums are the information paradox
\cite{page,hawkinginfo}, the holographic principle
\cite{thooft2,susskind} (for some developments see
\cite{bousso1,gaolemos1,gaolemos2,gaolemos3}), and last but not the
least the inside of a black hole and the problem of 
spacetimes singularities \cite{israel,penrosebook}.
All of these are problems in fundamental physics whose solutions
will help in a better understanding of the connections 
between quantum theory, statistical and information theory,  
and gravitation, and ultimately can lead us to 
the correct quantum theory of gravity.

\section*{Acknowledgments}

I thank the organizing committee of the Fifth International Workshop
on New Worlds in Astroparticle Physics, for providing a very
stimulating atmosphere in the gracious city of Faro, in a meeting,
held in January 2005, that opened up the celebrations of the first
World Year of Physics. I thank conversations with my students 
Gon\c calo Apr\'a Dias and Nuno Santos, as well as with Vitor Cardoso and
\'Oscar Dias. I also thank Observat\'orio Nacional do Rio de Janeiro
for hospitality.  This work was partially funded by FCT - Portugal 
through project POCTI/FNU/57552/2004.


\begin{thebibliography}{99}

\bibitem{lemosastroreview} J. P. S. Lemos, {\em astro-ph}/9612220 (1996). 

\bibitem{wheeler1} J. A. Wheeler, 
in {\em Relativity, groups an topology}, eds. C. and B. de Witt 
(Gordon and Breach, 1964), p. 315.

\bibitem{wheelerautobiog} J. A. Wheeler, K. Ford,
{\em Geons, black holes, and quantum foam: A life in physics}, 
(W. W. Norton \& Company, 2000). 

\bibitem{penrose} R. Penrose, {\em Nuov. Cimento} {\bf 1}, 
252 (1969).

\bibitem{hawking} S. W. Hawking, G. F. R. Ellis, 
{\em The large scale structure of space-time}, 
(Cambridge University Press, 1973). 

\bibitem{carterleshouches} B. Carter, in {\em Black holes, Les 
astres occlus}, ed. C. and B. de Witt 
(Gordon and Breach, 1973), p. 57. 

\bibitem{bekenstein0} J. D. Bekenstein, {\em Phys. Rev. D} 
{\bf 9}, 2333 (1973). 

\bibitem{hawking2}  S. W. Hawking, {\em Nature} {\bf 248}, 30 (1974); 
{\em Commun. Math. Phys.} {\bf 43}, 199 (1975). 

\bibitem{bekenstein1} J. D. Bekenstein, {\em gr-qc}/9710076 (1997).

\bibitem{mehra} J. Mehra, H. Rechenberg, 
{\em The historical development of quantum theory} (6 Volumes), 
(Springer, 2002).

\bibitem{witten} E. Witten, 
{\em Phys. Rev. D} {\bf 44}, 314 (1991).

\bibitem{horowitzstrominger}  G. T. Horowitz,  
A. Strominger, {\em Nucl. Phys.} {\bf B360}, 197 (1991). 

\bibitem{lemos1} J. P. S. Lemos, 
{\em Class. Quantum Grav.} {\bf12}, 1081 (1995).

\bibitem{2} J. P. S. Lemos, 
{\em Phys. Lett.}    {\bf B353}, {46} (1995). 

\bibitem{3} J. P. S. Lemos, V. T. Zanchin, 
{\em Phys. Rev. D} {\bf 54}, 3840 (1996).

\bibitem{4} J. P. S. Lemos, {\em gr-qc}/0011092 (2000).

\bibitem{bekensteinreview1} J. D. Bekenstein, {\em gr-qc}/9808028 (1998).

\bibitem{bekensteinreview2} J. D. Bekenstein, {\em gr-qc}/9409015 (1994).

\bibitem{fursaevreview}  D. V. Fursaev,
{\em Phys. Elem. Part. Nucl.} {\bf 36}, 81 (2005). 

\bibitem{pagereview} D. N. Page,  {\em New J. Phys.}
{\bf 7}, 203 (2005). 

\bibitem{padmanabhanreview} T. Padmanabhan, 
{\em Phys. Rep.} {\bf 406}, 49 (2005).

\bibitem{largeextradimensions} N. Arkani-Hamed, S. Dimopoulos,   
G. Dvali, {\em Phys. Lett.} {\bf B429}, 263 (1998).

\bibitem{mtw} C. W. Misner, K. S. Thorne, J. A. Wheeler, 
{\em Gravitation}, (Freeman 1973). 

\bibitem{kerr} R. Kerr, {\em Phys. Rev. Lett.} {\bf 11}, 237 
(1963).

\bibitem{lemosreview1997} J. P. S. Lemos,  {\em hep-th}/9701121 (1997).

\bibitem{zeldovich1} Ya. B. Zel'dovich, {\em Sov. Phys. JETP} {\bf 35}, 1085
(1972).

\bibitem{unruh} W. G. Unruh, {\em Phys. Rev. D} {\bf 10}, 3194 (1974).

\bibitem{bekenstein3} J. D. Bekenstein, M. Schiffer, 
{\em Phys. Rev. D} {\bf 58}, 064014 (1998).

\bibitem{Christod}  D. Christodoulou, {\em Phys. Rev. Lett.} {\bf 25},
1596 (1970).

\bibitem{HawkingPRL} S. W. Hawking, {\em Phys. Rev. Lett} {\bf 26}, 1344
(1971).

\bibitem{blackbomb} V. Cardoso, O. J. C. Dias, J. P. S. Lemos, S.
Yoshida, {\em Phys. Rev. D} {\bf 70}, 044039 (2004).

\bibitem{gulbenk2004} O. J. C. Dias, J. P. S. Lemos, ``Superradiant 
amplification in astrophysical black hole systems'', unpublished (2004). 

\bibitem{bardeenetal} J. M. Bardeen, B. Carter, S. W. Hawking,
{\em Commun. Math. Phys.} {\bf 31}, 161 (1973).

\bibitem{waldbook} R. Wald, {\em Quantum field theory in curved spacetime 
and black hole thermodynamics} (University of Chicago Press, 1994).

\bibitem{bekensteingeneralizedsecondlaw} J. D. Bekenstein, 
{\em Phys. Rev. D} {\bf 9}, 3292 (1974). 


\bibitem{bekenstein75} J. D. Bekenstein,  
{\em Phys. Rev. D} {\bf 12}, 3077 (1975).

\bibitem{frolovnovikov} V. Frolov, I. Novikov, 
{\em Phys. Rev. D} {\bf 48}, 4545 (1993). 

\bibitem{jacobson} T. Jacobson, {\em gr-qc}/9908031 (1999).

\bibitem{jacobsonmarolf} T. Jacobson, D. Marolf, 
C. Rovelli, {\em hep-th}/0501103 (2005).

\bibitem{sorkinlast} R. D. Sorkin, {\em hep-th}/0504037 
(2005).

\bibitem{gerlach} U. H. Gerlach, 
{\em Phys. Rev. D} {\bf 14}, 1479 (1976). 

\bibitem{zurek}  W. H. Zurek, K. S. Thorne, 
{\em Phys. Rev. Lett.} {\bf 54}, 2171 (1985). 

\bibitem{thooft} G. 't Hooft, 
{\em Nucl. Phys.} {\bf B256},727 (1985).

\bibitem{sorkin86} L. Bombelli, R. K. Koul, J. Lee, R. 
D. Sorkin, {\em Phys. Rev. D} {\bf 34}, 373 (1986). 

\bibitem{srednicki}  M. Srednicki, 
{\em Phys. Rev. Lett.} {\bf 71}, 666 (1993). 

\bibitem{adami} C. Adami, {\em quant-ph}/0405005 (2004).

\bibitem{sakharov} A. D. Sakharov, Sov. Phys. Dokl. {\bf 12}, 
1040 (1968); Gen. Rel. Grav. {\bf 32}, 365 (2000).

\bibitem{frolovfursaev} V. P. Frolov, D. V. Fursaev, 
{\em Phys. Rev. D} {\bf 56}, 2212 (1997). 

\bibitem{solodukhin}  S. N. Solodukhin, 
{\em Phys. Lett.} {\bf B454}, 213 (1999). 

\bibitem{carlip99} S. Carlip, 
{\em Class. Quant. Grav.} {\bf 16}, 3327 (1999).

\bibitem{carlip02} S. Carlip, {\em Phys. Rev. Lett.} 
{\bf 88}, 24130 (2002).

\bibitem{carlip05} S. Carlip, 
{\em Class. Quant. Grav.} {\bf 22}, 1303 (2005).

\bibitem{gdiaslemos} G. A. S. Dias, J. P. S. Lemos,
{\em Phys. Rev. D}, submitted (2005).

\bibitem{bekmukha}  J. D. Bekenstein, V. F. Mukhanov, 
{\em Phys. Lett.} {\bf B360}, 7 (1995). 

\bibitem{hod}  S. Hod, {\em Phys. Rev. Lett.} 
{\bf 81}, 4293 (1998). 

\bibitem{who?} H. P. Nollert, {\em Phys. Rev. D} 
{\bf 47}, 5253 (1993). 

\bibitem{motleneitzke} L. Motl, A. Neitzke, 
{\em Adv. Theor. Math. Phys.} {\bf 7}, 307 (2003). 

\bibitem{baez} J. Baez, in {\em gr-qc}/0303027 (2003).

\bibitem{cardosolemos} V. Cardoso, J. P. S. Lemos, S. Yoshida, 
{\em Phys. Rev. D} {\bf 69}, 044004 (2004). 

\bibitem{cardosothesis} V. Cardoso, 
{\em gr-qc}/0404093 (2004). 

\bibitem{york} J. W. York, 
{\em Phys. Rev. D}  {\bf 28}, 2929 (1983). 

\bibitem{loweetal}  N. Iizuka, 
D. Kabat, G. Lifschytz, D. A. Lowe, 
{\em Phys. Rev. D}  {\bf 67}, 124001 (2003). 

\bibitem{gibbonshawking} G. W. Gibbons, S. W. Hawking, 
{\em Phys. Rev. D} {\bf 15}, 2738 (1977).

\bibitem{teitel1996} C. Teitelboim, 
{\em Phys. Rev. D} {\bf 51}, 4315 (1995).

\bibitem{wald} V. Iyer, R. M. Wald, 
{\em Phys. Rev. D} {\bf 50}, 846 (1994).  

\bibitem{hawkinghunter1999}  S. W. Hawking, C. J. Hunter, 
{\em Phys. Rev. D} {\bf 59}, 044025 (1999).

\bibitem{garfinkleetal}  D. Garfinkle, S. B. Giddings, 
A. Strominger, 
{\em Phys. Rev. D} {\bf 49}, 958 (1994).

\bibitem{diaslemos0} O. J. C. Dias, J. P. S. Lemos, 
{\em Phys. Rev. D} {\bf 69}, 084006 (2004). 

\bibitem{diaslemos1} O. J. C. Dias, 
{\em Phys. Rev. D} {\bf 70}, 024007 (2004). 

\bibitem{diaslemos2}  O. J. C. Dias, J. P. S. Lemos, 
{\em Phys. Rev. D} {\bf 70}, 124023 (2004).

\bibitem{diasthesis} O. J. C. Dias, {\em hep-th}/0410294 (2004).

\bibitem{kol} B. Kol, {\em hep-th}/0411240 (2004).

\bibitem{horowitzpolchinski}  G. T. Horowitz, J. Polchinski, 
{\em Phys. Rev. D}  {\bf 55}, 6189 (1997). 

\bibitem{mathur} S. M. Mathur, 
{\em Fortschr. Phys.}  {\bf 53}, 793 (2005). 

\bibitem{stromingervafa}  A. Strominger, C. Vafa, 
{\em Phys. Lett.} {\bf B379}, 99 (1996).

\bibitem{cardy} J. A. Cardy, {\em Nucl. Phys.} {\bf B270}, 186 (1986).

\bibitem{strominger}  A. Strominger, 
{\em JHEP}  {\bf 9802}, 009 (1998).

\bibitem{brownhenneaux} J. D. Brown, M. Henneaux, 
{\em Comm. Math. Phys.} {\bf 104}, 207 (1986).

\bibitem{maldacenaAdsCFTconjecture} J. M. Maldacena, 
{\em Adv. Theor. Math. Phys.} {\bf 2}, 231 (1998).

\bibitem{page}  D. N. Page, {\em hep-th}/9305040 (1993).

\bibitem{hawkinginfo} S. W. Hawking, {\em hep-th}/0507171 (2005).

\bibitem{thooft2} G. 't Hooft, {\em gr-qc}/9310026 (1993).

\bibitem{susskind} L. Susskind, {\em J. Math. Phys.} 
{\bf 36}, 6377 (1995). 

\bibitem{bousso1} R. Bousso, {\em Rev. Mod. Phys.} {\bf 74}, 825 (2004).

\bibitem{gaolemos1} S. Gao, J. P. S. Lemos,  
{\em JHEP} {\bf 0404}, 017 (2004). 

\bibitem{gaolemos2}  S. Gao, J. P. S. Lemos,  
{\em  Phys. Rev. D} {\bf 71}, 084010 (2005).

\bibitem{gaolemos3} S. Gao, J. P. S. Lemos,  
{\em in the Proceedings of the
fifth international workshop on new worlds in astroparticle physics}, 
eds. Ana M. Mour\~ao et al., (World Scientific 2006), p. 272. 

\bibitem{israel} W. Israel, in  {\it Black holes and relativistic 
stars}, ed. R. Wald (University of Chicago Press, 1998), p. 137.

\bibitem{penrosebook} R. Penrose, 
{\em The road to reality: A complete guide to the laws of the Universe}, 
(Knopf, 2005). 



\end{thebibliography}
\end{document}